\documentclass{jetpl_arXiv11114931}   
\twocolumn

\newcommand{\beq}{\begin{equation}}
\newcommand{\eeq}{\end{equation}}
\newcommand{\beqa}{\begin{eqnarray}}
\newcommand{\eeqa}{\end{eqnarray}}
\newcommand{\bsubeqs}{\begin{subequations}}
\newcommand{\esubeqs}{\end{subequations}}

\title{\vspace*{4mm}
Spontaneously broken Lorentz invariance from the dynamics of a heavy sterile neutrino}

\rtitle{Spontaneously broken Lorentz invariance}

\sodtitle{}

\author{F.R. Klinkhamer $^{\#}$
\/\thanks{\:e-mail: frans.klinkhamer@kit.edu}}

\rauthor{F.R. Klinkhamer}

\sodauthor{Klinkhamer}

\address{$^{\#}$ Institute for Theoretical Physics,
Karlsruhe Institute of Technology (KIT), 76128 Karlsruhe, Germany}


\abstract{A relativistic theory for neutrino superluminality is
presented (in principle, the same mechanism applies also to other fermions).
The theory involves the standard-model particles and one additional heavy
sterile neutrino with an energy-scale close to or above the electroweak one,
all particles propagating in the usual $3+1$
spacetime dimensions. Lorentz violation results from spontaneous symmetry
breaking in the sterile-neutrino sector. The theory tries, as far as
possible, to be consistent with the existing experimental data from neutrino
physics and to keep the number of assumptions minimal. There are clear
experimental predictions which can be tested.
\vspace*{-4.5mm}}

\begin{document}

\maketitle

\section{Introduction}
\label{sec:Introduction}

The time-of-flight results of the MINOS~\cite{MINOS2007} and
OPERA~\cite{OPERA2011,OPERA2011-press} long-baseline neutrino
experiments are, for the moment, inconclusive as to the
existence of  a superluminal muon-neutrino velocity.
At the same level of accuracy,
the ICARUS Collaboration~\cite{ICARUS2012}
reports finding a standard (luminal) value for the
muon-neutrino velocity.
Awaiting further experimental results, theory may benefit
by performing a ``fire drill,'' best made
as realistic and as difficult as possible.

Let us assume, for the sake of argument, that there is indeed neutrino
superluminality: $v/c-1 >0$ with $c$ the speed of light in vacuum.
And take the superluminality of a $10\;\text{GeV}$ neutrino
to be at the $10^{-6}$ level,
that is, relatively large but a factor $25$ below the original OPERA
claim~\cite{OPERA2011} and compatible with the
result of ICARUS~\cite{ICARUS2012}.

The challenge, then, is to find a theoretical explanation that is both
consistent and convincing. The qualification ``consistent'' refers to all
other experimental facts established by elementary particle physics to
date. The qualification ``convincing'' refers to
the desire to introduce as few additional assumptions as
possible~\cite{Endnote:post-OPERA-papers}.

Before presenting one possible theoretical explanation,
let us give a list of experimental facts:
\begin{enumerate}
\item[(i)]
The qualitative (low-statistics) ICARUS bound~\cite{ICARUS2012}
on the time-of-flight muon-neutrino velocity at an energy of order
$10\;\text{GeV}$:
$|v_{\,\nu_\mu}(10\;\text{GeV})\,- c\,|/c \lesssim 5 \times 10^{-6}$.
Our theoretical ``fire drill''
takes the fiducial value
$(v_{\,\nu_\mu}(10\;\text{GeV})\,- c\,)/c =10^{-6}$.
\item[(ii)]
The supernova SN1987a bound on the time-of-flight electron-antineutrino
velocity~\cite{SN1987a,Longo1987}:
$|v_{\,\overline{\nu}_e}(10\;\text{MeV})\,- c\,|/c
\lesssim 10^{-9}$. (Bunching may even suggest a tighter bound
at the $10^{-13}$ level.)
\item[(iii)]
The existence of coherent
mass-difference neutrino oscillations, which requires
nearly equal maximum velocity of the three known flavors
($f=e,\,\mu,\,\tau$) of neutrinos~\cite{ColemanGlashow1998}.
\item[(iv)]
The absence of significant energy losses~\cite{CohenGlashow2011}
from the vacuum-Cherenkov-type process
$\nu_\mu\to \nu_\mu+Z^{0} \to \nu_\mu+e^{-}+e^{+}$ at tree-level,
in particular, for the CERN--GranSasso  (CNGS) neutrinos
detected by OPERA and ICARUS.
\item[(v)]
The negligible leakage~\cite{Giudice-etal2011} of Lorentz violation
from the neutrino sector into the charged-lepton sector
by quantum effects (e.g., loop corrections to the electron propagator).
\end{enumerate}
There may be other experimental facts to explain, but the
five listed above [with the relatively large fiducial value of point (i)
at the $10^{-6}$ level] suffice for the moment as they
are already difficult enough to understand.

Points (i) and (iv) are especially difficult to reconcile, that is,
having superluminality but no Cherenkov-type energy losses.
If we do not wish to introduce new light particles
(such as light sterile neutrinos with or without extra
spacetime dimensions~\cite{Giudice-etal2011,%
HannestadSloth2011,Nicolaidis2011,Winter2011,Klinkhamer2011}),
the simplest approach may be to seek a \emph{reduction} of the
vacuum-Cherenkov rate, instead of the complete absence of the
process. It has, indeed, been pointed out that
a strong energy dependence of the neutrino velocity excess
may lead to a significant reduction of the
vacuum-Cherenkov rate~\cite{MohantyRao2011}.
The details of the calculation of Ref.~~\cite{MohantyRao2011} still
need to be verified, but, as will be explained later on,
it is also possible to understand
the reduction factor analytically~\cite{KaufholdKlinkhamer2005,%
KaufholdKlinkhamer2007}.
It, therefore, appears to be attractive phenomenologically to
consider an energy-dependent modification of the
neutrino dispersion relation which leads to superluminality.

\newpage
The fundamental problem, however, is to give an  explanation
for the superluminal maximum velocity of these neutrinos. A relatively
simple explanation relies on spontaneous breaking of Lorentz
invariance~\cite{KlinkhamerVolovik2011,NojiriOdintsov2011}.
In the following, we give one possible realization
of such a 4-dimensional superluminal-neutrino model.
Different from the previous discussion in
Ref.~\cite{KlinkhamerVolovik2011}, special attention is paid to
the issue of gauge invariance.

\section{Theory}
\label{sec:Theory}

Let us start at the phenomenological level and then work
our way down to the underlying theory.
In terms of the 3-momentum norm  $p\equiv |\mathbf{p}|$,
the following model dispersion relations of the three neutrino
mass eigenstates ($n=1,\,2,\,3$) are considered:
\beqa\label{eq:mod-disp-sbli-mass-basis}
\big[E_{n}(p)\big]^{2} &\sim&   c^{\,2}\,p^{2}
             +\big(m_{n}\,c^{\,2}\big)^{2}
             +c^{-4}\,(b^{0})^{8}\;M^{-6}\;p^{8}\,,
\eeqa
with an identical eighth-order term for all three dispersion relations
in terms of
a dimensionless constant $b^{0}\in\mathbb{R}$ and a mass scale $M$,
both resulting from a fermion condensate to be discussed shortly.
The single eighth-order term in \eqref{eq:mod-disp-sbli-mass-basis} is
only an example and is supposed to hold for sufficiently low values of
$E_{n}/Mc^{2}$,
as will be discussed further in Sec.~\ref{sec:Discussion}.

Henceforth, we set $c=\hbar=1$ and use the Minkowski metric,
\beq\label{eq:metric}
g_{\alpha\beta}(x) = \eta_{\alpha\beta} \equiv
\text{diag}(1,\,-1,\,-1,\,-1)\,,
\eeq
together with the standard notation $\partial_{\alpha}$
for the partial derivative $\partial/\partial x^{\alpha}$
(operating to the right, as usual).
The theory without Lorentz violation is defined by
the Lagrange density $\mathcal{L}_\text{SM}$
of the standard model (SM)~\cite{Veltman1994},
to which are added three neutrino mass terms ($n=1,\,2,\,3$),
\beq\label{eq:L-mn}
\mathcal{L}_{m_{n},\,\text{kin}}=-m_{n}\,\overline{\psi}_{n}\;\psi_{n}
+ \overline{\psi}_{nR}\:i\gamma^\alpha\partial_\alpha\:\psi_{nR}
\,,
\eeq
in terms of neutrino Dirac fields $\psi_{n}(x)$.
Also included in \eqref{eq:L-mn} are the kinetic terms for
the noninteracting right-handed components, as these terms
are absent in $\mathcal{L}_\text{SM}$.

The modified dispersion relations
\eqref{eq:mod-disp-sbli-mass-basis} now come from
additional Lorentz-violating (LV) Lagrange-density terms
which take the form of standard mass terms
$\mathcal{L}_{m_{n}}$ with derivative operators
inserted between the Dirac spinors~\cite{KlinkhamerVolovik2011}.
Specifically, the relevant higher-derivative term is
\bsubeqs\label{eq:quadraticbInteraction-b-timelike}
\beq\label{eq:quadraticbInteraction}
\mathcal{L}_{\nu-\text{LV}}=-M^{-3}\;\sum_{n=1}^{3}\;
\overline{\psi}_{n}\:[b^\alpha\,\partial_\alpha ]^4\;\psi_{n} \,,
\eeq
where the background vector is assumed to be purely timelike,
\beq\label{eq:b-timelike}
(b^\alpha)=(b^{0},\,0,\,0,\,0)\,.
\eeq
\esubeqs
The dispersion relation \eqref{eq:mod-disp-sbli-mass-basis}
then results for small enough values of the mass and energy,
$\max[m_{n},\,E] \ll M$ for $|b^{0}|\sim 1$.
Terms of the type \eqref{eq:quadraticbInteraction}
may also provide for superluminality of fermionic particles
other than neutrinos, but here the focus is on neutrinos.

Again following Ref.~\cite{KlinkhamerVolovik2011},
the dimensionless background vector $b^\alpha$
arises as a fermion condensate,
\beq\label{eq:b-condensate}
b^\alpha=M^{-4}\;\eta^{\alpha\beta}\;
<\overline{\chi}_{S}\,(-i\,\partial_\beta)\,\chi_{S}>\,,
\eeq
where $\chi_{S}(x)$ is the Dirac field of a heavy sterile neutrino
with mass scale $M\gg \text{max}(|m_1|,\,|m_2|,\,|m_3|)$.
Dynamically, the condensate \eqref{eq:b-condensate} may come
from the following multi-fermion interaction term:
\begin{subequations}\label{eq:S-self-interaction}
\begin{eqnarray}\label{eq:S-self-interaction-pot}
\mathcal{L}_{S,\,\text{int}} &=&
-\lambda\;M^{4}\; \Big( X - B^{2} \Big)^{2}\,,
\\[2mm]
\label{eq:S-self-interaction-def-X}
X&\equiv&  M^{-8}\;\eta^{\alpha\beta}\,
\big[\overline{\chi}_{S}\,(-i\,\partial_\alpha)\, \chi_{S}\big]
\big[\overline{\chi}_{S}\,(-i\,\partial_\beta)\,  \chi_{S}\big] ,
\end{eqnarray}
\end{subequations}
with real coupling constants $\lambda$ and $B$
(alternatively, $\lambda$ can be considered to be a
Lagrange multiplier; cf. Ref.~\cite{NojiriOdintsov2011}).
Suitable boundary conditions
and small symmetry-breaking perturbations (later taken to zero,
the standard procedure for the study of spontaneous symmetry breaking)
pick out a condensate vector
\eqref{eq:b-condensate} which is purely timelike \eqref{eq:b-timelike}
with time-component
\beq\label{eq:b-condensate-timelike}
b^{0}=\pm\, B\,.
\eeq
Remark that, with the chosen signature \eqref{eq:metric},
the interaction term \eqref{eq:S-self-interaction}
is only able to select a nonvanishing timelike condensate vector,
not a spacelike one.

An appropriate higher-derivative interaction term
produces the effective
momentum-dependent mass terms \eqref{eq:quadraticbInteraction}
for the propagators of the light (active) neutrinos.
Using the notation of Ref.~\cite{ChengLi1984}
and correcting a typo in its Eq.~(13.69),
one possible gauge-invariant relativistic interaction term reads
\beqa\label{eq:L-nu-S-int}
\hspace*{-0mm}&&
\mathcal{L}_{\nu S,\,\text{int}}=
\Big(M^{19}\,v/\sqrt{2}\,\Big)^{-1}\,
\nonumber\\
\hspace*{-0mm}&&
\sum_{f=e,\,\mu,\,\tau}
\Big[\overline{L}_{f}\cdot \widetilde{\Phi}\;
[\big(\overline{\chi}_{S}\,\partial^\alpha\chi_{S}\big)\,
\partial_\alpha ]^4\,\nu_{R,f}
+ \text{H.c.}\Big]\,,
\eeqa
with the usual left-handed lepton isodoublets $L_{f}(x)$
and Higgs isodoublet $\Phi(x)$ of the standard model, together with
three (ultra-)heavy right-handed sterile neutrinos $\nu_{R,f}(x)$
[which may or may not be related to the right-handed projection
of our previous field $\chi_{S}(x)$].
The Higgs vacuum expectation value $v$ is defined by
$<\Phi^\dagger\cdot\Phi>\,\equiv v^{2}/2$.
With $\widetilde{\Phi}\equiv (i\tau_2) \cdot \Phi^\ast$ and
$\overline{L}_{f}\,\cdot <\widetilde{\Phi}>$ $=$
$\overline{\nu}_{L,f}\;v/\sqrt{2}\,$,
expression \eqref{eq:L-nu-S-int}
corresponds to a sum of identical Dirac-mass-type terms,
$\sum_{f}\,[\overline{\nu}_{L,f}\,\mathcal{M}\,\nu_{R,f}+
\overline{\nu}_{R,f}\,\mathcal{M}^\dagger\,\nu_{L,f}]\,$,
which ultimately gives \eqref{eq:quadraticbInteraction}
by use of \eqref{eq:b-condensate}.

All in all, the underlying relativistic theory
of the model dispersion relations \eqref{eq:mod-disp-sbli-mass-basis}
has the following Lagrange density:%
\begin{subequations}\label{eq:L-total-LI-SSB-Skin}
\beqa\label{eq:L-total}
\mathcal{L} &=&
\mathcal{L}_\text{LI}+\mathcal{L}_\text{SSB}\,,\\[1mm]
\label{eq:L-LI-alt}
\mathcal{L}_\text{LI} &=&
\mathcal{L}_\text{SM}+\sum_{n}\;\mathcal{L}_{m_{n},\,\text{kin}}^{\,\prime}
+\mathcal{L}_{M,\,\text{kin}}\,,
\\[1mm]
\label{eq:L-SSB-alt}
\mathcal{L}_\text{SSB} &=&
\mathcal{L}_{S,\,\text{int}}+
\mathcal{L}_{\nu S,\,\text{int}}\,,
\\[1mm]
\label{eq:L-Skin-alt}
\mathcal{L}_{M,\,\text{kin}} &=&
\overline{\chi}_{S}\,\big(i\,\gamma^\alpha\partial_\alpha-M\big)\,\chi_{S}\,,
\eeqa
\end{subequations}
where \eqref{eq:L-Skin-alt} is the standard kinetic Dirac term for
the heavy sterile neutrino and the other
parts of the standard Lorentz-invariant (LI) term \eqref{eq:L-LI-alt}
have been discussed in the second paragraph of this section.
The prime on the light-neutrino mass term in
\eqref{eq:L-LI-alt} indicates that
it is gauge invariant and written in terms of the interacting SM
fields~\cite{ChengLi1984}, having additional  derivative terms
for the noninteracting fields
as discussed below \eqref{eq:L-mn}.

The spontaneous symmetry breaking (SSB) term \eqref{eq:L-SSB-alt}
involves the interaction terms \eqref{eq:S-self-interaction}
and \eqref{eq:L-nu-S-int}, which are manifestly gauge-invariant
because the heavy-neutrino field $\chi_{S}$ is a gauge singlet
(normal derivatives suffice).
As these interaction terms are non-renormalizable, it can be
expected that a different, more fundamental theory applies for
center-of-mass scattering energies \mbox{$\sqrt{s} \gtrsim M$.}

This completes the presentation of one particular relativistic
theory with spontaneous breaking of Lorentz invariance,
giving rise to superluminal neutrinos propagating in the usual
$3+1$ spacetime dimensions.

\section{Discussion}
\label{sec:Discussion}

Return to the phenomenological level \eqref{eq:mod-disp-sbli-mass-basis}
of our relativistic theory \eqref{eq:L-total-LI-SSB-Skin}. Then,
the neutrino group velocity $v_{n,\,\text{gr}}$
has a relative superluminality
proportional to $B^{8}\,(E_n)^{6}/M^{6}$, in terms of the heavy-neutrino scale
$M$ and the coupling constant $B$ entering the heavy-neutrino
self-interaction term \eqref{eq:S-self-interaction} and
giving the dispersion-relation parameter $b^{0}$
by \eqref{eq:b-condensate-timelike}.
A hypothetical superluminality at the $10^{-6}$ level
of a $10\;\text{GeV}$ muon-neutrino
implies a mass scale $M \sim 10^{2}\;\text{GeV}$ for $|b^{0}|=|B|\sim 1$.

Let us now go through the list of experimental facts
from Sec.~\ref{sec:Introduction}.
The ICARUS bound from point (i) is obviously satisfied.
The supernova bound from point (ii) is also satisfied because,
according to \eqref{eq:mod-disp-sbli-mass-basis},
the modification of the group velocity $v_{n,\,\text{gr}}$
has a sixth-order energy dependence, so that
a decrease of the energy
from $10\;\text{GeV}$ to $10\;\text{MeV}$
reduces superluminality effects by an additional factor of $10^{-18}$.
[As to the actual energy dependence of $v_{n,\,\text{gr}}(E)$,
it has been argued~\cite{Giudice-etal2011} that
these functions must peak at  $E \sim 10\;\text{GeV}$ or
reach a plateau for $E \gtrsim 10\;\text{GeV}$~\cite{Endnote:energy-dependence}.
In our case, this may require further higher-derivative terms
of the type \eqref{eq:quadraticbInteraction}.
Needless to say, the discussion of the present article is at the
level of an ``existence proof,'' leaving aside all questions of
naturalness.]
Point (iii) holds because the theory is constructed to give
an identical eighth-order term for all three dispersion
relations \eqref{eq:mod-disp-sbli-mass-basis}.
The last two points, (iv) and (v), are more subtle.

The vacuum-Cherenkov-type process of point (iv) is not forbidden
but has a significantly reduced rate~\cite{MohantyRao2011}.
In fact, a heuristic argument based on the concept of
effective-mass-squares~\cite{ColemanGlashow1998,CohenGlashow2011,KaufholdKlinkhamer2005}
gives an extra factor
of order $\big(1/\sqrt{7}\,\big)^{5} \approx 10^{-2}$
for the tree-level decay rate
$\Gamma(\nu_\mu\to\nu_\mu+e^{-}+e^{+})
 \propto (G_F)^{2}\,(m_\text{eff})^{5}$
in our theory \eqref{eq:L-total-LI-SSB-Skin}
compared to the rate in the
theory with an identical Lorentz-violating $p^{2}$ term in the three
neutrino dispersion relations.
Incidentally, the Lorentz-violating vacuum-Cherenkov-type process
is similar to the standard Cherenkov-radiation process
but not identical, as discussed in Secs. III and IV of
Ref.~\cite{KaufholdKlinkhamer2007}.

Point (v) regarding the Lorentz-violation leakage into the charged-lepton
sector remains problematic~\cite{Giudice-etal2011}.
It is all the more important to find a good theoretical explanation for
point (v), as the experimental bounds of the maximum velocities
of certain standard-model particles are extremely tight~\cite{Altschul2007}.

In addition, it remains for us to better understand the dynamic origin of
the particular Lorentz-violating term \eqref{eq:quadraticbInteraction},
because the required higher-derivative interaction terms were introduced
in a more or less \textit{ad hoc} fashion.
See, e.g., Refs.~\cite{Bjorken1963,Jenkins2004,Kostelecky2003,Klinkhamer2012}
for further discussion on spontaneous breaking of Lorentz invariance.

Leaving these fundamental properties of the theory aside, it
is evident that the superluminal-neutrino
model \eqref{eq:mod-disp-sbli-mass-basis}
as it stands leads to direct predictions for experiment.
With the CNGS setup, for example,
a narrow symmetric pulse of nearly mono-energetic
muon-neutrinos produced at CERN (cf. Sec.~9 of Ref.~\cite{OPERA2011})
would give a nearly equal pulse shape for the
final muon-neutrinos to be detected by OPERA and ICARUS.
The final pulse profile would be not exactly identical to the
initial one, as a percent or so of the muon-neutrinos would have
lost energy due to vacuum-Cherenkov
electron-positron-pair emission~\cite{MohantyRao2011},
which would have reduced the speed of these neutrinos somewhat
according to \eqref{eq:mod-disp-sbli-mass-basis}.
The superluminal-sterile-neutrino
hypothesis~\cite{Giudice-etal2011,%
HannestadSloth2011,Nicolaidis2011,Winter2011,Klinkhamer2011} typically
predicts a substantial broadening of the pulse profile.

Assuming neutrino superluminality to exist in the first place,
detailed measurements of the final pulse profile may
thus provide information about the underlying mechanism.
The main point of this article, however, has been to show that
it is possible to obtain
spontaneously broken Lorentz invariance from a four-dimensional
gauge-invariant relativistic theory if there is a
heavy sterile neutrino with appropriate higher-derivative
self-interactions.

\section*{ACKNOWLEDGMENTS}\noindent
It is a pleasure to thank G.E.~Volovik, D.~Zeppenfeld, and the referee
for helpful remarks on an earlier version of this article.


\end{document}